\documentclass[aps,prb,twocolumn,superscriptaddress,10pt,article,nofootinbib,showpacs,longbibliography,nofootinbib]{revtex4-2}
\usepackage{blindtext}
\usepackage{lipsum}
\usepackage{graphics}
\usepackage{amsmath}
\usepackage{graphicx}
\usepackage{graphics}
\usepackage{amssymb}
\usepackage{verbatim}
\usepackage{physics}
\usepackage{float}
\usepackage[normalem]{ulem}
\usepackage[colorlinks, citecolor=blue]{hyperref}
\usepackage[dvipsnames]{xcolor}
\usepackage{bbm}

\def \beq{\begin{eqnarray}}
\def \eeq{\end{eqnarray}}

\begin{document}

\title{Reply to Comment on ``Discrete Time Crystals: Rigidity Criticality and Realizations''}
\author{Norman Y. Yao}
\affiliation{Department of Physics, University of California, Berkeley, CA 94720 USA}
\author{Andrew C. Potter}
\affiliation{Department of Physics, University of British Columbia, Vancouver, BC V6T-1Z4 Canada}
\author{Ionut-Dragos Potirniche}
\affiliation{Department of Physics, University of California, Berkeley, CA 94720 USA}
\author{Ashvin Vishwanath}
\affiliation{Department of Physics, Harvard University, Cambridge, MA 02138 USA}

\begin{abstract}
This is a reply to the comment from Khemani, Moessner and Sondhi~(KMS)~[arXiv:2109.00551] on our manuscript  [Phys. Rev. Lett. 118, 030401 (2017)]. 
The main new claim in KMS is that the short-ranged model does not support an MBL DTC phase.  We show that, even for the parameter values they consider and the system sizes they study, the claim is an artifact of an unusual choice of range for the crucial plots.
Conducting a standard finite-size scaling analysis
on the same data 
strongly suggests that the system is in fact a many-body localized (MBL) discrete time crystal (DTC). 
Furthermore, we have carried out additional simulations at larger scales, and provide an analytic argument, which fully support the conclusions of our original paper. 
We also show that the effect of boundary conditions, described as essential by KMS, is exactly what one would expect, with boundary effects decreasing with increasing system size. 
The other points in KMS are either a rehashing of points already in the literature (for the long-ranged model) 
or are refuted by a proper finite-size scaling analysis.

\end{abstract}

\maketitle
In our paper \cite{yao2017discrete}, henceforth YPPV, we proposed a model for a many-body localized (MBL) discrete time crystal (DTC), whose main ingredients were sufficiently simple for quantum simulators existing at the time of publication to approximately realize.

Given that it has been  nearly five years since YPPV was first released, it may be helpful to  recall the  status of the field at that time.
Although it was understood that an MBL DTC phase could be realized with relatively few ingredients, the proposed toy models needed fields, couplings and disorder strengths, that were not easily realizable in existing experimental platforms. 
A key question YPPV addressed was, can discrete time crystals be realized while being mindful of the constraints imposed by experimental feasibility at the time? 
Despite being less ideal than the toy models, the models introduced in YPPV were demonstrated to realize time crystalline behavior. 
This helped convert the field of time crystals into an experimental one, with experiments first performed in trapped ions~\cite{zhang2017observation} and NV centers~\cite{choi2017observation} and then in many other platforms.

We begin by summarizing the objections raised in the comment by Khemani, Moessner, and Sondhi (KMS)~\cite{khemani2021comment}.
The most significant objection is the claim that the phase diagram associated with the Floquet unitary [Eqn.~(1) in YPPV~\cite{yao2017discrete}]:
\begin{equation}
    U = e^{-i\sum_i \left(J^z_i \sigma^z_i \sigma^z_{i+1} + B^z_i \sigma^z_i\right)}
    e^{-i\left(\frac{\pi}{2}-\epsilon\right)\sum_i \sigma_i^x}
    \label{eq:model}
\end{equation}
does not contain an MBL DTC.  Eq.~\ref{eq:model} represents a short-range interacting spin-1/2 chain of length $L$, where $\vec{\sigma}$ are Pauli operators, $B^z_i \in [0, W]$ is a random longitudinal field, and $J^z_i \in [J_z - \delta J_z, J_z + \delta J_z]$.

KMS arrive at their conclusion numerically by examining three quantities: (i) spin autocorrelations under open boundary conditions (OBC), (ii) spin autocorrelations under periodic boundary conditions (PBC), and (iii) mutual information under open boundary conditions.
KMS speculate on a different scenario to explain the observed oscillations, based on a thermalized phase in proximity to an ordered ground state. 

 KMS claim that our observed numerical signatures of MBL DTC behavior, including long-time correlations in the infinite-temperature spin autocorrelator, and mutual information (MI) between distant spins are artifacts of either finite-size or boundary condition effects. 
Here, we refute these claims through a careful analysis of finite-size scaling, parameter sweeps, and the influence of boundary conditions.

\vspace{2mm}

Our central contentions are the following:
\begin{itemize}
\item We show analytically that there is a regime of MBL DTC at sufficiently small $\epsilon$ for any non-zero $J_z,\delta J_z$.
\item KMS claim that the long-time plateau associated with period-doubled oscillations in $\sigma^z(t)$ exhibits a finite size drift indicating that it vanishes in the infinite system size ($L\rightarrow \infty$) limit.
This is done without any quantitative extrapolation but rather by visually inspecting data plotted on unusual axes.
Instead, we show via a systematic finite-size scaling analysis that the period-doubled oscillation amplitude converges to a non-zero value as $L\rightarrow \infty$ indicative of an MBL DTC. 
We further confirm our conclusion by simulating larger system sizes. 
We show this at the same point in the phase diagram as KMS studied and for two other parameters. 

\item KMS claim that the period-doubled oscillations and long-range mutual information signatures of the MBL DTC observed in YPPV are artifacts of using open-boundary conditions (OBC) due to the presence of almost-strong edge modes~\cite{kemp2017long,else2017prethermalSZM}. 
To address this point, we re-examine these observables under periodic boundary conditions. We demonstrate that the MBL DTC signatures persist in the absence of edges.
\end{itemize}

\begin{figure*}
\includegraphics[width=7.0in]{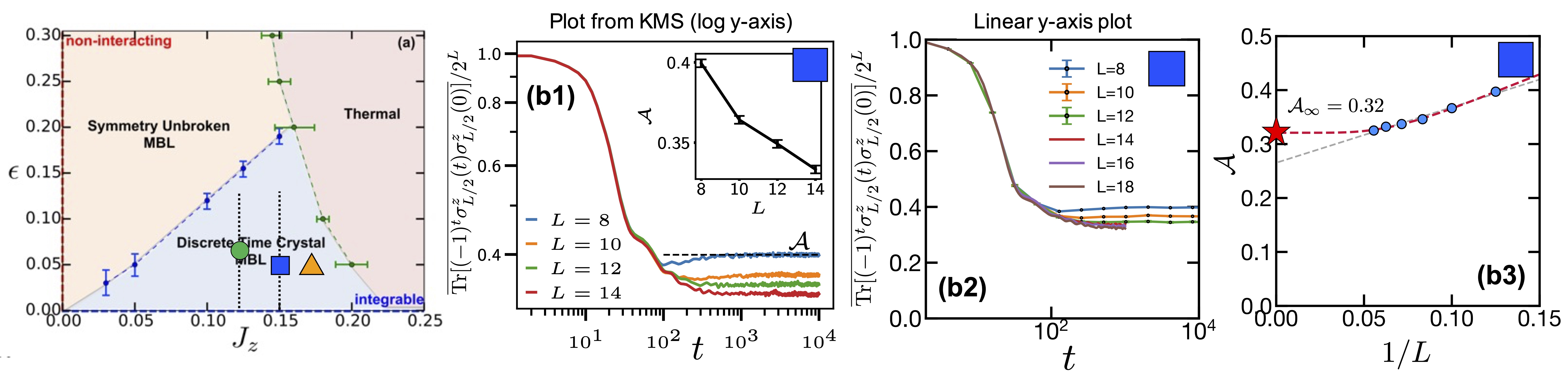}
\caption{%
(a1) Phase diagram from \cite{yao2017discrete}.~The three markers indicate the  parameters in the phase diagram where the infinite temperature autocorrelation function, $A_{L/2}(t)$, is computed. 
The two dashed black lines indicate the parameter sweep associated with $\mathcal{I}$ in Figure~\ref{fig:MI_PBC}.
(b1) Depicts $A_{L/2}(t)$ as a function of time for $J_z=0.15$, $\epsilon=0.05$. For clarity, only even periods are shown.  
(b2) Depicts the analogous plot from KMS at the same parameter value as in b1. Note that the y-axis is on a logarithmic scale that ends at approximately 0.3. (inset) depicts $\mathcal{A}$ as extracted by KMS. Where system sizes overlap, our data agrees with KMS.
(b3) Depicts the late-time plateau value, $\mathcal{A}$, extracted from b1 as a function of $1/L$.
$\mathcal{A}(L)$ is fit using the functional form $\mathcal{A}_\infty + be^{-L/2\xi}$ (red dashed curve).
A finite intercept (indicated by the red star) indicates the presence of stable period-doubling with a finite amplitude as $L\rightarrow \infty$.
The extracted values are $\{\mathcal{A}_\infty,b,\xi\} = \{0.32, 0.63, 1.9 \}$. The gray dashed line indicates a linear  fit that also shows a non-zero intercept.
}
\label{fig:logaxis}
\end{figure*}

{\bf Existence of an MBL DTC phase at small pulse-detuning}---Before embarking on a detailed analysis of numerical results, we first argue analytically that Eq.~\ref{eq:model}  hosts an MBL DTC phase at small pulse-detuning, $\epsilon$. 

At $\epsilon=0$, Eq.~\ref{eq:model} can be written in the form:
\begin{equation}
    U=W^\dagger e^{-iH_F}X W,
\label{eq:udecomp}
\end{equation}
where $W|_{\epsilon=0}=\mathbbm{1}$, $X=\prod_i \sigma_i^x$ is a perfect spin-flip, and the Floquet Hamiltonian is a zero correlation length disordered Ising spin-glass: 
$H_F=\sum_i J_i \sigma_i^z\sigma_{i+1}^z$.
The combination of the spin-glass eigenstates of $H_F$ and the perfect spin-flip action of $X$ yield ideal DTC behavior. Crucially, at  $\epsilon=0$, for any non-zero value of the parameters $J_z,\delta J_z$, the spectrum of $H_F$ is labeled by a complete set of strictly-local integrals of motion (LIOMs) that are symmetric under a $\mathbb{Z}_2$ symmetry generated by $X$. 

By now, there is extensive numerical and analytic evidence~\cite{abanin2019colloquium}, and in some cases rigorous proofs~\cite{imbrie2016many}, that adding generic short-range perturbations below a critical threshold to such a model 
results in a finite region of MBL phase.
Specifically for Floquet systems, there is strong analytic and numerical evidence that, below a critical strength, perturbing strictly-localized models will preserve the decomposition [Eq.~\ref{eq:udecomp}], but where $W$ becomes a finite-depth local unitary, and $H_F$  becomes a generic MBL Hamiltonian (with a non-zero but finite localization length) that obeys an emergent symmetry generated by $X$~\cite{harper2020topology}.

Moving away from $\epsilon =0$ will generate transverse fields at $O(\epsilon)$ and integrability-breaking terms at $O(\epsilon^2)$. The precise form of $H_F$ at $\epsilon \neq 0$ can be approximated by standard high-frequency expansion techniques~\cite{else2017prethermal} (see also Eq.~4 of KMS)~\footnote{This HF expansion misses rare local resonance effects that need to be calculated non-perturbatively, but which have been shown through a variety of techniques~\cite{abanin2019colloquium} not to disrupt MBL below a critical interaction strength.}.

\begin{figure*}
\includegraphics[width=7.0in]{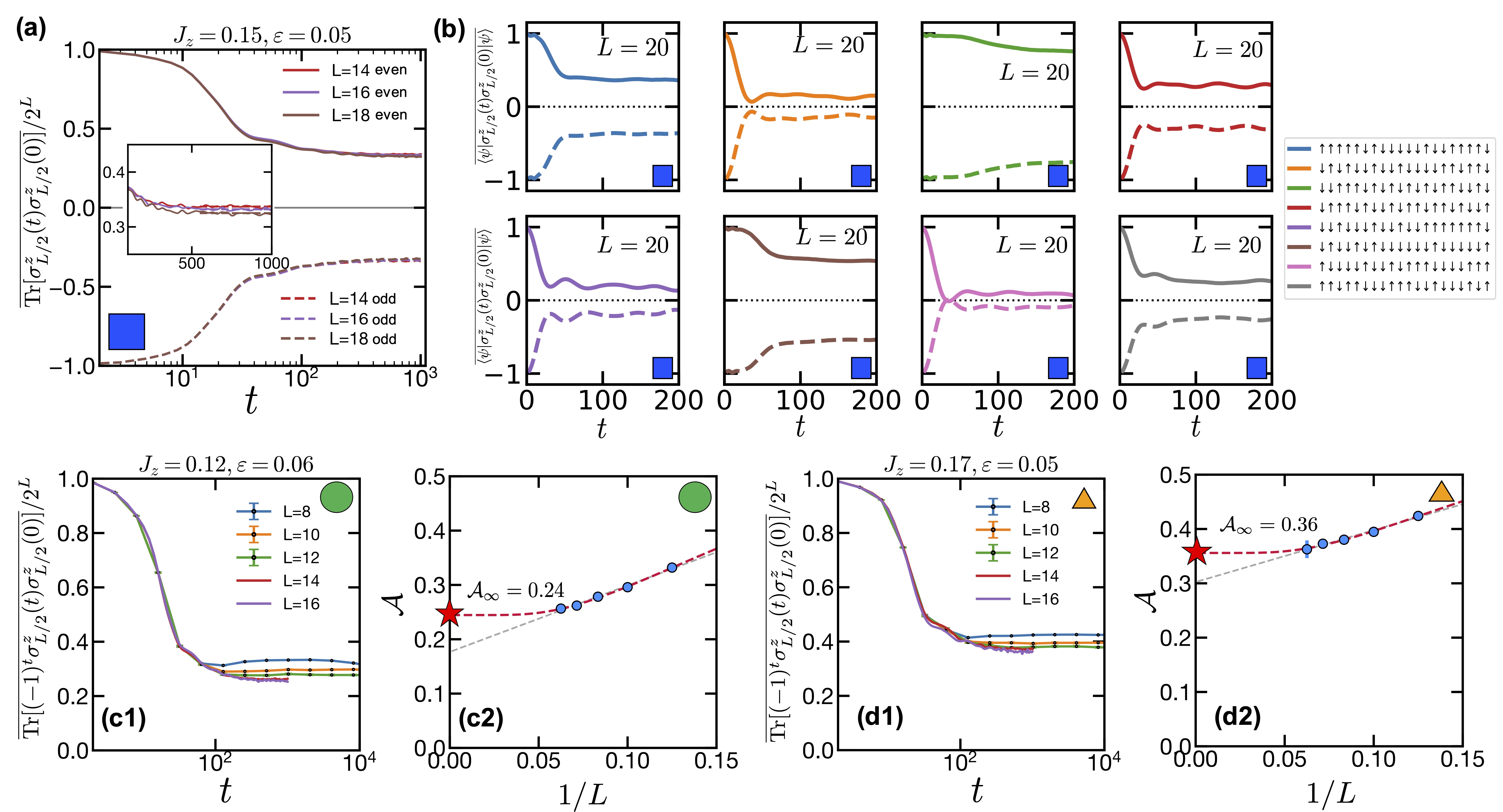}
\caption{%
(a) Infinite temperature spin autocorrelator for $J_z=0.15$, $\epsilon = 0.05$ [same parameter as Fig.~\ref{fig:logaxis}(b)] showing both even and odd cycles.  (inset) shows a zoomed in version where one can distinguish a small difference in the late-time plateau value for $L=14,16,18$. 
(b) Dynamics of the spin autocorrelation function at site $L/2$ starting from eight random initial product states. Dynamics are computed for system size $L=20$ using Krylov methods and averaged over 100 disorder realizations. Even periods are shown as solid lines, while odd periods are shown as dashed lines. 
(c,d) Depict the same analysis as Fig~\ref{fig:logaxis} for two other points in the phase diagram of Fig~\ref{fig:logaxis}a.
(c1,c2) Depicts the same analysis [as performed in Fig.~\ref{fig:logaxis}] of the late-time plateau  of the infinite temperature autocorrelator  for the parameter $J_z=0.12$, $\epsilon = 0.06$ (green circle in Fig.~\ref{fig:logaxis}a).
A finite value of the late-time plateau as $L\rightarrow \infty$ is obtained through a  scaling analysis. 
In particular, the extracted values are  $\{\mathcal{A}_\infty,b,\xi\} = \{0.24, 0.66, 2.0 \}$.
(d1,d2) Depicts the same analysis for the parameter $J_z=0.17$, $\epsilon = 0.05$ (gold triangle in Fig.~\ref{fig:logaxis}a).
The extracted values are $\{\mathcal{A}_\infty,b,\xi\} = \{0.36, 0.53, 1.9 \}$.
All data for $L=8,10,12$ are obtained via exact diagonalization. Data for $L=14,16,18$ are obtained via Kyrlov subspace methods. 
The number of disorder averages varies depending on system size: 40k for $L=8$, 10k for  $L=10$, 8k for $L=12$. For the Krylov data: $J_z=0.12$ (6k for $L=14$, 5k for $L=16$), $J_z=0.15$ (11k for $L=14$, 13k for $L=16$, 5k for $L=18$), $J_z=0.17$ (9k for $L=14$, 1.5k for $L=16$).
}
\label{fig:Ainf}
\end{figure*}

Thus, in the $\epsilon <\epsilon_c \lesssim \delta J_z$ regime, $H_F$ will  be in an MBL spin-glass phase which spontaneously breaks  the emergent Ising symmetry. 
The $X$ pulses in $U$ will then lead to period-two oscillations of this emergent symmetry-breaking pattern, i.e.~produce an MBL DTC. 
This  establishes the existence of an MBL DTC phase in the model [Eq.~\ref{eq:model}] for any $J_z,\delta J_z \neq 0$ (this regime was explicitly explored in YPPV) refuting the claim of KMS. 

The only remaining question is then to map out the extent of this MBL DTC phase.
Determining the phase boundaries cannot  be accomplished by examining  individual points in the phase diagram as in KMS~\cite{khemani2021comment}, but rather requires a careful analysis of finite-size scaling and parameter dependence --- a task to which we now turn.

\begin{figure*}
\includegraphics[width=7.in]{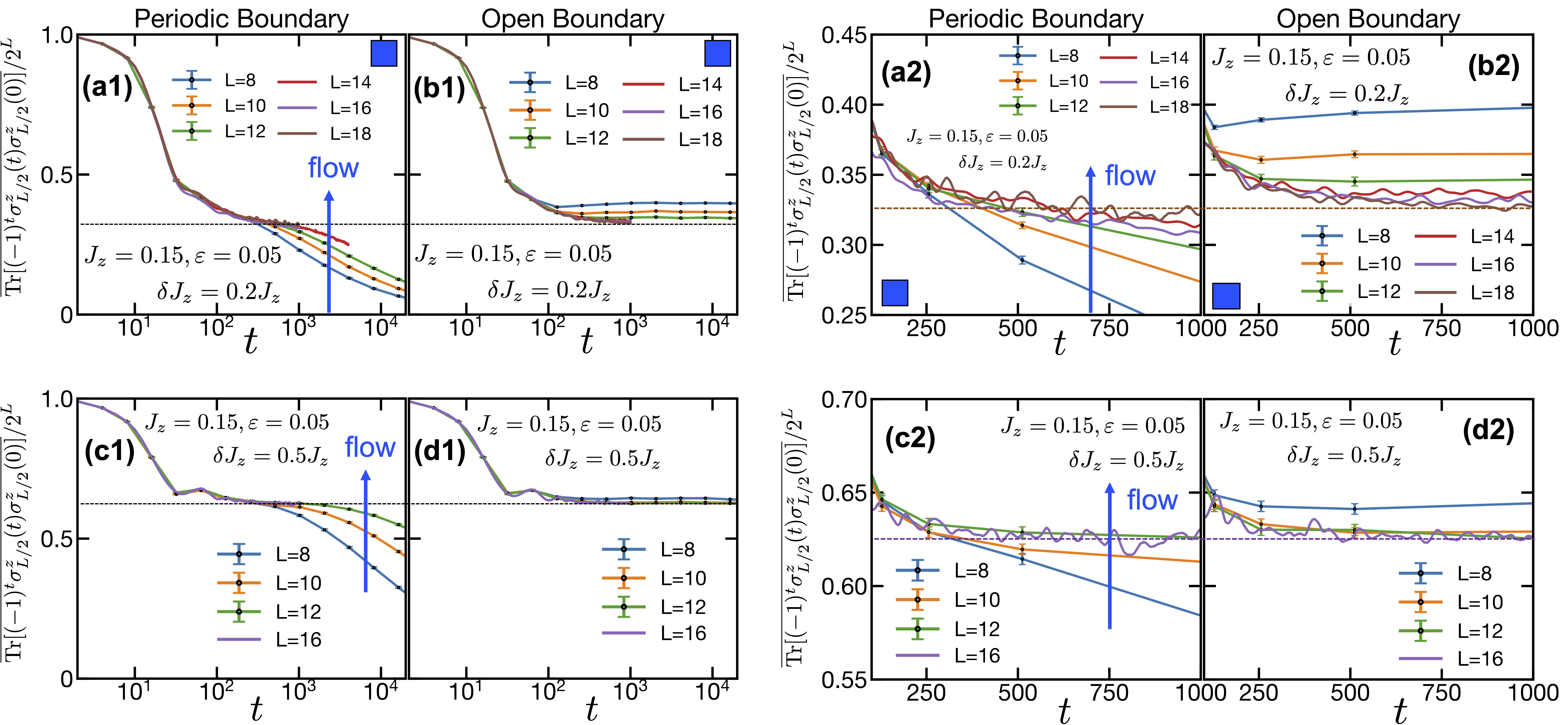}
\caption{%
Comparison of the infinite temperature autocorrelation function, $A_{L/2}(t)$, for periodic versus open boundary conditions. 
(a1, b1) Depict $A_{L/2}(t)$ for the parameter $J_z=0.15$, $\epsilon=0.05$, $\delta J_z = 0.2 J_z$ with periodic (a1) and open (b1) boundary conditions. 
The periodic boundary condition case exhibits an increasing flow as a function of system size, consistent with flow toward the same plateau as found in the open boundary condition case.
(a2, b2) Show zoomed in versions of (a1, b1), demonstrating that at $L=18$, the late-time plateau value at $t=10^3$ is nearly identical for the two different boundary conditions.
(c1, d1) Depict $A_{L/2}(t)$ for the parameter $J_z=0.15$, $\epsilon=0.05$, $\delta J_z = 0.5 J_z$ (not originally simulated in YPPV) with periodic (c1) and open (d1) boundary conditions. 
The flow in the periodic boundary condition case is even more dramatic and saturation  toward the same plateau as found in the open boundary condition case occurs for smaller system sizes. 
(c2, d2) Show zoomed in versions of (c1, d1), demonstrating that at $L=16$, the late-time plateau value at $t=10^3$ is nearly identical for the two different boundary conditions.
\label{fig:PBC_flow}
}
\end{figure*}

{\bf Presence of stable period doubling}---Stable period-doubling in local observable dynamics is the hallmark of a discrete time crystal, and is reflected by a long-time plateau in the (state-averaged) spin autocorrelations:
\begin{equation}
    A_i(t) = \frac{1}{2^L}\Tr [(-1)^t \sigma^z_i(t) \sigma^z_i(0)].
\end{equation}
In Fig.~1(b) of their comment, KMS show that the plateau amplitude can depend on the initial state, suggesting that ``long-lived oscillations'' only arise ``from special initial states (such as polarized states).''
In Fig.~\ref{fig:Ainf}(b), we demonstrate that this is not the case. %
We simulate the Floquet dynamics of eight totally random initial states (far from the polarized regime) and observe long-lived oscillations with amplitudes consistent with the infinite temperature average (to be discussed shortly). 
In fact, it is challenging to find states that do not exhibit such long-lived oscillations with a finite plateau amplitude. 

Nevertheless, finding specific ``bad-actor'' states~\cite{khemani2021comment} that exhibit low values of the plateau amplitude is to be expected for small system sizes due to the strongly disordered nature of the model.
Indeed, when one moves away from the exactly-solvable DTC limit ($\epsilon=0$), one must rely upon statistical measures that average over states. 
The real question is if the MBL DTC order survives after this averaging?
KMS acknowledge this point and ``present data for infinite temperature autocorrelators'', stating that  since such a correlator ``averages over initial states, it is less sensitive to outlier states  such as the polarized one."
KMS  acknowledge the presence of  a late-time plateau in the DTC regime identified by YPPV, but, based on observing a slight decrease of the plateau with system size for the parameter point examined [Fig.~\ref{fig:logaxis}(b2)], conclude that this plateau will vanish for $L\rightarrow \infty$, stating: ``We find that the short-range model does display a plateau as appropriate to an an MBL DTC. However, this is a finite size effect, with the amplitude $\mathcal{A}$ showing a clear decrease with increasing system size.''

We emphasize that any exact diagonalization (ED) simulations on modest system sizes will exhibit some finite size flow, and that the $L\rightarrow \infty$ behavior cannot be determined by simply  the ``sign'' of the finite size corrections. 
Rather, one must perform a careful, quantitative extrapolation to the $L\rightarrow \infty$ limit, to which we now turn.

\paragraph{Finite-Size Scaling:} Using the parameters from YPPV, $W = 2\pi$ and $\delta J_z = 0.2 J_z$, in Fig.~\ref{fig:logaxis}(b2) and Fig.~\ref{fig:Ainf}(a,c1,d1), we plot the infinite temperature autocorrelation function for the spin at site $L/2$ using open boundary conditions. 
In addition to the parameter chosen by KMS [blue square, Fig.~\ref{fig:logaxis}(a)], we also show data for two other parameters as indicated via the markers in the phase diagram in Fig.~\ref{fig:logaxis}(a). 
For the specific parameter chosen by KMS ($J_z = 0.15$,  $\epsilon = 0.05$), our data (at system sizes which overlap) agrees with KMS [see comparison between Fig.~\ref{fig:logaxis}(b1) and (b2)].
KMS chose to plot $A_{L/2}(t)$ using an interesting logarithmic y-axis (beginning at approximately 0.3), reproduced in Fig.~\ref{fig:logaxis}(b1),  which exaggerates the finite-size dependence of the late-time plateau value, $\mathcal{A}$. 
Thus, in Fig.~\ref{fig:logaxis}(b2), we plot the same infinite-temperature autocorrelation function  (consistent with their data), on a  linear scale from zero to one, the natural range for this order parameter. 
By employing a Krylov subspace technique, we are able to push the data to system sizes up to $L=18$ and times $t\sim 10^3$.
The plateau value appears to saturate to a non-zero value with increasing system size. 
However, one cannot conclude by visual inspection alone what the $L\rightarrow \infty$ behavior will be, as such inspection can be fooled by data plotting choices. 
To this end, we undertake a careful  finite-size scaling analysis.

For each parameter [Fig.~\ref{fig:logaxis}(a)], we extract the late-time plateau value, $\mathcal{A}$,  of $A_{L/2}(t)$ as a function of system size.
In Fig.~\ref{fig:logaxis}(b3), we plot $\mathcal{A}$ versus $1/L$.
Note that at larger system sizes, the $1/L$ corrections begin to  decelerate and the curvature is upwards.
In an MBL phase, one expects exponentially decaying correlations for local observables, and we fit $\mathcal{A}(L)$ to the functional form $\mathcal{A}_\infty + be^{-L/2\xi}$  [red dashed curve,  Fig.~\ref{fig:logaxis}(b3)].
The data follow this form  and  extrapolate to a non-zero value, $\mathcal{A}_\infty = 0.32$, as $L\rightarrow \infty$, indicating stable period doubling.
A more naive and stringent extrapolation, using the functional form $\mathcal{A}(L) =a+b/L$  [gray dashed curve,  Fig.~\ref{fig:logaxis}(b3)] also extrapolates to a finite value for $\mathcal{A}_\infty$.
This directly contradicts the claim that the plateau is a finite-size artifact made by KMS.
Identical conclusions are reached for two other parameters in the YPPV MBL-DTC phase diagram, $J_z=0.12$, $\epsilon=0.06$ [see Fig.~\ref{fig:Ainf}(c1,c2)] and $J_z=0.17$, $\epsilon=0.05$ [see Fig.~\ref{fig:Ainf}(d1,d2)].

\paragraph{Influence of boundary conditions:} 
KMS also criticize the use of open boundary conditions  due to the presence of ``almost strong" edge modes in the small $\epsilon$ regime~\cite{kemp2017long,else2017prethermalSZM}. 
Specifically, they claim that for periodic boundary conditions, $A_{L/2}(t)$ shows ``only a steady decay with time with no visible plateau.'' 

We now demonstrate that this conclusion is based on simulating a limited range of small system sizes and does not reflect the $L\rightarrow \infty$ behavior.
Indeed, on general theoretical grounds, as $L\rightarrow \infty$, boundary conditions cannot influence the physics of local bulk observables.

\begin{figure*}
\includegraphics[width=7.0in]{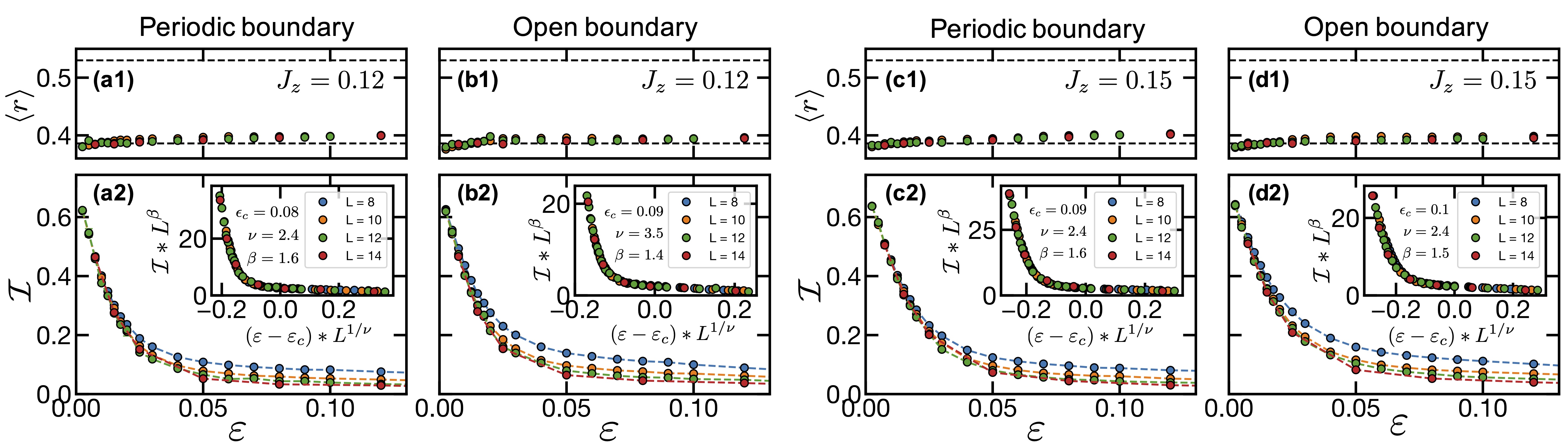}
\caption{Depicts the mutual information, $\mathcal{I}_{1,L/2}$, between the spin at site $1$ and site $L/2$ using periodic boundary conditions (a,c panels), where there are no edges, and the  mutual information, $\mathcal{I}_{L/2-\lfloor L/4\rfloor,L/2+\lfloor L/4\rfloor}$, using open boundary conditions (b,d panels). 
Parameter sweeps as a function of $\epsilon$ are taken for fixed $J_z=0.12$ and $J_z=0.15$.
(a1, b1, c1, d1) Depict the level statistics, $\langle r \rangle$-ratio, as a function of $\epsilon$.
The bottom dashed line indicated the Poisson value consistent with MBL, while the top dashed line indicates the thermal value.
(a2, b2, c2, d2) Depicts the MI as a function of pulse-detuning. 
The MI exhibits the same qualitative behavior for both OBC and PBC.
(inset) Depicts a critical scaling collapse.
}
\label{fig:MI_PBC}
\end{figure*}

For $J_z = 0.15$, $\epsilon=0.05$ (the parameter chosen in KMS and in Fig.~\ref{fig:logaxis}(b), Fig.~\ref{fig:Ainf}(a,b) above), we plot $A_{L/2}(t)$ for both periodic [Fig.~\ref{fig:PBC_flow}(a1)] and open boundary conditions [Fig.~\ref{fig:PBC_flow}(b1)]  side-by-side.  
As we have already established, the open boundary condition case exhibits flow toward a finite value of the late-time plateau, $\mathcal{A}$.
As depicted in Fig.~\ref{fig:PBC_flow}(a1), for periodic boundary conditions, $A_{L/2}(t)$ indeed exhibits a decay for small system sizes.
However, we find that the finite size flow in $A_{L/2}(t)$ is \emph{opposite} --- in particular, $A_{L/2}(t)$ increases  with increasing $L$.
Moreover, Figures~\ref{fig:PBC_flow}(a2) and (b2) depict a zoomed in comparison, which shows that at larger system sizes (e.g.~$L=18$ at times $\sim 10^3$), $A_{L/2}(t)$ exhibits a plateau value which is consistent for both PBC and OBC.
This directly refutes the claim of KMS that the plateau associated with long-lived DTC oscillations is an artifact of boundary conditions.
Additional evidence for this type of behavior---namely, that both PBC and OBC converge toward the same plateau value as system sizes increase (as expected on general grounds)---is provided by Figures~\ref{fig:PBC_flow}(c1,c2,d1,d2), which simulate $A_{L/2}(t)$ for $J_z = 0.15$, $\epsilon=0.05$, $\delta J_z=0.5 J_z$.

\paragraph{Collective resonances and avalanches:} 
Another point made by KMS is that  finite-size numerical evidence for an MBL DTC phase could give way to ``avalanche instabilities'' at very large system sizes. Recent research~\cite{de2017stability,crowley2020avalanche}, suggests that collective resonances
arising in system sizes beyond the reach of existing numerical techniques
decrease the critical interaction strength associated with delocalization.
At present, while there is qualitative evidence and arguments for avalanches, there are no rigorous tools to quantify their effects. 
KMS quote an estimate for the critical single-particle localization length, $\xi^*_{sp}\approx 2.88$;  we caution that this number is model-dependent and non-universal.
We note that, since arbitrarily small localization lengths can be achieved for small enough $\epsilon$, even avalanches cannot qualitatively change the phase diagram of YPPV.

{\bf Mutual Information}---The dynamical signatures discussed in the previous section definitively establish the presence of an MBL-DTC phase in Eq.~\ref{eq:model} and invalidate KMS's main contentions.
Due to their importance for numerical studies, we next address KMS's critique of Floquet eigenstate signatures such as the long-range mutual information (MI)~\cite{else2016floquet}.
Long-range mutual information (MI) between distant pairs of spins provides a theoretically important and basis-independent measure of MBL DTC order~\cite{else2016floquet}. 
 KMS argue that MI is a flawed measure because it varies from state-to-state across the Floquet spectrum, and has many states in which MI is  lower than its maximal value of $\log 2$.
 In addition, they point out that in open chains, the MI between end spins can include effects from   the presence of ``almost strong" Majorana edge modes.
 Here, we instead show that MI indeed provides robust signatures of MBL-DTC behavior.

\paragraph{State-to-state variations:}
To start, we note that state-to-state variations in the  two-spin MI (with values less than $\log 2$) are expected in an MBL DTC for $\epsilon\neq 0$.
In particular, there is a non-zero probability that either of the spins are pinned to a locally strong field and do not participate in the extensive ``DTC" cluster; in addition, the local conserved quantities generically are spread out over multiple spins. 
Since these effects vary across states and disorder realizations, in an MBL-DTC one expects a distribution of the two-spin MI about a non-zero average and with non-zero variance.
In particular, there is a non-vanishing probability of finding MI arbitrarily close to zero even as $L\rightarrow \infty$.
The mutual information behaves like an ``order parameter'', which scales continuously to zero at the DTC-melting transition. 
Therefore, simply noting the presence of individual states with low MI, as in KMS, is not sufficient to conclude the absence of a DTC phase. Instead, one must rely upon statistical measures that average over states and the key question is whether the state- and disorder- averaged MI remains finite in the thermodynamic limit.
Thus, 
a systematic finite-size analysis, such as the critical scaling collapse conducted in YPPV, is crucial.

\paragraph{Edge modes and boundary conditions:} KMS further contend that the ``presence of Majorana edge modes in the ordered phase''  implies that our  mutual information diagnostic ``merely reflects the edge physics of an ordered transverse field Ising model.'' We demonstrate that this is incorrect by computing the MI under periodic boundary conditions, where we observe DTC signatures without the presence of edges. 

 While the MI between spins $1$ and $L$ in an open system, as presented in YPPV,  includes contributions from both edge-state physics and bulk DTC physics, the latter is readily isolated as follows. 
We compute the MI, $\mathcal{I}_{i,j}$, between two bulk spins at sites $i=L/2-\lfloor L/4\rfloor$ and $j=L/2 + \lfloor L/4\rfloor$ with OBC and at sites $i=1$ and $j=L/2$ with PBC (Fig.~\ref{fig:MI_PBC}).
For both OBC and PBC, we compute the MI along two distinct cuts (sweeping $\epsilon$) of the phase diagram [black dashed lines in Fig.~\ref{fig:logaxis}(a)]. 
We  observe  quantitative differences between $\mathcal{I}_{L/2-\lfloor L/4\rfloor,L/2 + \lfloor L/4\rfloor}$ and  $\mathcal{I}_{1,L}$ from the OBC data in YPPV, indicating the latter includes the effect of edges. 
However, in contrast to KMS's main claim, the MI for both OBC and PBC (where there are no edges) depicted in Fig.~\ref{fig:MI_PBC} is consistent with an extended regime of MBL-DTC, showing saturating $L$ dependence to a non-zero value for both types of boundary conditions.

In addition, we perform a finite-size scaling collapse with $L$ based on the expected scaling form for a continuous DTC-melting transition [inset, Fig.~\ref{fig:MI_PBC}(a2,b2,c2,d2)].
This analysis provides evidence for an MBL DTC phase that persists up to a critical pulse-detuning,  beyond which a continuous DTC-melting transition occurs.
This is consistent with the observation of stable period-doubled MBL DTC oscillations at the parameter studied in KMS as well as the other parameters marked in Fig.~\ref{fig:logaxis}(a).
Of course, the usual caveats about extracting critical exponents from small-scale exact diagonalization data apply.

{\bf Long-range model and trapped ion experiments}---Finally, KMS point out that the ``ab initio" model of the trapped ion experiments (Eq.~\ref{eq:model}, but with power-law ranged Ising interactions $J_z \sim 1/r^{p}$ with $p\approx 1.5$) is not an MBL DTC. 
While the tension between long-range interactions and MBL was already referred to in YPPV~\cite{yao2014many,de2017stability}, the full complex interplay between long-range interactions, many-body localization,  avalanches and 1D prethermal time crystals was the subject of multiple later works and remains an active area of research~\cite{potirniche2019exploration,crowley2020avalanche,machado2020long}.
These points have already been acknowledged and extensively discussed in recent reviews~\cite{khemani2019brief,else2020discrete}.  

We emphasize, however,  that the classical period-doubling of zero-dimensional oscillators which KMS implies is related to prethermal DTCs, is qualitatively distinct from prethermal DTCs, which are a many-body phenomenon that is stabilized by interactions and occurs for a finite parameter range.

{\bf Summary}
To summarize, a straightforward analytic argument clearly establishes the existence of an MBL DTC phase in the model of Eq.~\ref{eq:model} for small $\epsilon$.
In YPPV and in the present response, we have conducted a careful, systematic numerical study across a wide range of parameters and system sizes.
Through our analysis, we have presented strong numerical evidence against KMS's conclusions for our model defined in Eq.~\ref{eq:model}.

\emph{Acknowledgements}---We thank Dominic Else, Chetan Nayak, Joel Moore, Chris Monroe, and Romain Vasseur for insightful discussions and a critical reading of this reply.

\bibliography{vk_response.bib}

\end{document}